\documentclass[prl,twocolumn,footinbib,showpacs]{revtex4}
\usepackage{graphicx}
\usepackage{hyperref}
\hypersetup{colorlinks,citecolor=blue}

\begin{document}


\newcommand{\change}[1]{%
	\ensuremath{\clubsuit\!\triangleright}#1%
	\ensuremath{\triangleleft\!\clubsuit}%
}
\newcommand{\workon}[1]{%
	---\makebox[0pt]{$\triangleright$}$\bullet$#1%
	\makebox[0pt]{$\triangleleft$}$\bullet$---%
}



\title{Experimental Realization of an Optical One-Way Barrier for Neutral Atoms}
\author{Jeremy J. Thorn, Elizabeth A. Schoene, Tao Li, and Daniel A. Steck}
\affiliation{Oregon Center for Optics and Department of Physics, 1274 University
of Oregon, Eugene, OR 97403-1274}


\begin{abstract}
We demonstrate an asymmetric optical potential barrier for ultracold $^{87}$Rb
atoms using laser light tuned near the D$_2$ optical transition.
Such a one-way barrier, where atoms incident on one side are transmitted
but reflected from the other, is a realization of Maxwell's demon and has
important implications for cooling atoms and molecules not amenable to
standard laser-cooling techniques.
In our experiment, atoms are confined to a far-detuned dipole trap
consisting of a single focused Gaussian beam, which is divided near the
focus by the barrier.
The one-way barrier consists of two focused laser beams oriented almost
normal to the dipole-trap axis.
The first beam is tuned to present either a potential well or barrier,
depending on the state of the incident atoms.
On the reflecting side of the barrier, the second beam optically pumps the
atoms to the reflecting (barrier) state, thus producing the asymmetry.
\end{abstract}


\maketitle


One of the main goals in the field of atom optics is the development of new
tools to govern the motional states of atoms.
Recently, just such a tool for controlling atomic motion has been proposed:
the optical ``one-way barrier'' or ``atom diode''
\cite{Raizen05,Ruschhaupt04,Dudarev05,Kim05,Ruschhaupt06,Ruschhaupt06b,
	Ruschhaupt06c,Ruschhaupt07,Price07,Ruschhaupt08}.
The one-way barrier is an \emph{asymmetric} optical potential in that atoms
``see'' a different potential depending on the side of incidence.

The one-way barrier is particularly intriguing as a physical realization of
Maxwell's demon \cite{Maxwell71,Bennett87,Scully07,Ruschhaupt06c}.
In this thought-experiment, the demon manipulates a trapdoor in a wall
dividing a container of gas.
The demon could use the trapdoor as a one-way barrier to reduce the space
occupied by the atoms without performing any work, in an apparent violation
of the second law of thermodynamics.
This ``paradox'' has a number of resolutions \cite{Bennett87,Scully07}, but
the key issue is that the demon must perform measurements on the
system, gaining the information needed to perform feedback via the
trapdoor.
The entropy decrease of the atoms is balanced by an \emph{increase} in
the entropy of the demon's memory due to the accumulated information.
This cannot continue indefinitely in a cyclic process
unless the demon's memory is reset, or ``erased.''
The erasure requires work, in accordance with the second law.
In our experiment, the erasure occurs via the spontaneous scattering
of a barrier photon, which acts as an effective position measurement on
each atom and carries away the required entropy.

Clearly, then, the one-way barrier may be used to compress the phase-space
density of an ensemble of atoms, and thus to cool atomic samples
\cite{Raizen05}.
Laser-cooling techniques are now well-established \cite{Metcalf99}, but
generally rely on the existence of a cycling optical transition: atoms
falling into ``dark states'' decouple from the cooling lasers.
However, cooling techniques based on one-way barriers
\cite{Raizen05,Dudarev05,Ruschhaupt06c,Price07,Ruschhaupt08,Price08} rely
minimally on spontaneous scattering to produce the desired dissipation: in
principle, they can work by scattering only a \emph{single} photon.
In this regime, the problem of dark states is rendered moot, and in
principle the realization of one-way barriers paves the way for laser
cooling of atoms and molecules, vastly broadening the scope of laser
cooling \cite{Raizen05} without requiring optical cavities
\cite{Horak97,Chan03}.
Indeed, ``single-photon cooling'' has recently been demonstrated
for alkali atoms \cite{Price08}, using a mechanism closely related to the
one-way barrier here.
In addition to cooling, optical one-way barriers are likely to be generally
useful for manipulating atoms, particularly in guiding potentials as on
atom chips.

In the interest of implementing and optimizing cooling, it is critical to
study in detail the dynamics of atoms in the vicinity of the barrier.
In this Letter, we experimentally investigate a realization of an optical
one-way barrier for atoms confined to a dipole trap.
We have observed the dynamics of atoms crossing the barrier, and we
demonstrate the accumulation of atoms predominantly on one side of a
dipole-trap ``container.''



\begin{figure}[tb]
  \begin{center}
  \includegraphics[scale=0.8]{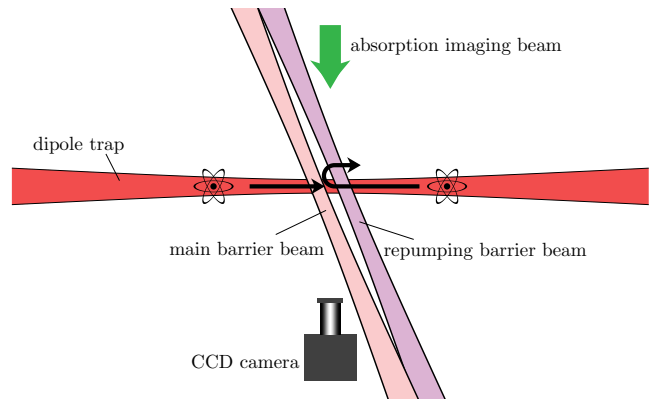}
  \end{center}
  \vspace{-5mm}
  \caption
        {(Color online.)
        Schematic diagram of the optical setup, showing the
        dipole trap, barrier beams, and imaging system.
      \label{fig:opticalsetup}}
\end{figure}

\begin{figure}[tb]
  \begin{center}
  \includegraphics[scale=0.86]{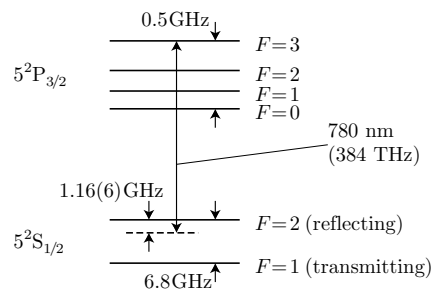}
  \end{center}
  \vspace{-5mm}
  \caption
        {Relevant atomic levels of $^{87}$Rb on the D$_2$ cooling
        transition (not to scale), shown with the off-resonant coupling of the
        main barrier laser.
        Atoms with $F=1$ see this field as
        red-detuned, while $F=2$ atoms see a blue detuning.
        The second barrier beam is resonant with the usual
        $F=1\rightarrow F'=2$ repumping transition.
	\label{fig:hfs}}
%
%
\end{figure}

The main feature of our experimental setup (Fig.~\ref{fig:opticalsetup}) is
the optical one-way barrier, which consists of two focused, Gaussian laser
beams.
The main barrier beam is tuned between the $F=1 \rightarrow F'$ and $F=2
\rightarrow F'$ $^{87}$Rb hyperfine transitions, as shown in
Fig.~\ref{fig:hfs}.
The detunings $\Delta:=\omega-\omega_0$ of the laser frequency $\omega$
with respect to the atomic resonances $\omega_0$ thus have opposite signs
for the two hyperfine ground levels.
As the optical dipole potential due to the laser field is proportional to
$I/\Delta$, where $I$ is the local laser intensity, the main barrier beam
presents a potential \emph{well} to atoms in the $F=1$ ground state, and a
potential \emph{barrier} to atoms in the $F=2$ ground state.
If we start with all atoms in the $F=1$ state, then they can pass unimpeded
through the barrier beam.
On one side of the barrier beam we have a second beam tuned to the $F=1
\rightarrow F'=2$ repump transition.
Atoms on that side of the barrier beam are thus optically pumped into the
$F=2$ ground state.
These atoms now see the main barrier beam as a potential barrier.
The repumping barrier beam thus defines the reflective side of the one-way
barrier, the side where the atoms are effectively trapped.

We start with $^{87}$Rb atoms in a standard six-beam magneto-optic trap (MOT)
\cite{Metcalf99}, loaded from a cold atomic beam in ultra-high vacuum
($\lesssim 10^{-10} \ \mathrm{torr}$).  After a secondary,
polarization-gradient-cooling stage, with increased detuning and reduced
intensity of the trapping light, we have about $2 \times 10^5$ atoms at about
$30 \ \mathrm{\mu K}$.

The atoms are then loaded into an optical dipole trap.
Our dipole-trap laser is a $1090 \ \mathrm{nm}$ (multiple longitudinal
mode) fiber laser which emits a collimated, Gaussian beam, with an
operating power of $9.3(5) \ \mathrm{W}$, which we focus to a $31.0(5) \
\mathrm{\mu m}$ waist ($1/e^2$ intensity radius), implying a $2.8 \
\mathrm{mm}$ Rayleigh length.
This beam produces a nearly conservative potential well, with a maximum
potential depth of $k_{_\mathrm{B}} \cdot 1.0 \ \mathrm{mK}$ for $^{87}$Rb
atoms in either hyperfine ground level, and a maximum scattering rate of
only $2 \ \mathrm{s^{-1}}$.
Near the trap center, the longitudinal and transverse harmonic frequencies
are $25 \ \mathrm{Hz}$ and $3.1 \ \mathrm{kHz}$, respectively.
Due to the large separation of the oscillation frequencies, we can regard
the atomic motion to be effectively one-dimensional along the dipole-beam
axis.
We typically start the atoms in the anharmonic region of the trap, which is
why the atoms dephase and have a different period ($48 \ \mathrm{ms}$) than
the harmonic frequency suggests.
The $1/e$ lifetime of atoms in the dipole trap by itself is $20
\ \mathrm{s}$.

We load the dipole trap from the MOT for $5 \ \mathrm{ms}$, trapping
about $3 \times 10^4$ atoms at $\sim\!\! 100 \ \mathrm{\mu K}$.
Longer load times can trap more atoms, but they spread throughout the trap.
We used a relatively short loading time, sacrificing atom number to resolve
the atomic dynamics with localized initial conditions.
For the data presented here, we loaded the atoms 
$0.95(5) \ \mathrm{mm}$ to either side of the focus by shifting the MOT with a
magnetic bias field.
Once loaded, the atoms in the dipole trap can be pumped into the $F=1$
ground state by extinguishing the MOT repump beam, and leaving the MOT
lasers (red-detuned by $70 \ \mathrm{MHz}$) on for another $7 \ \mathrm{ms}$.
We can also pump to the $F=2$ ground state using only the repump light.

\begin{figure*}[tb]
  \begin{center}
  \includegraphics[scale=1]{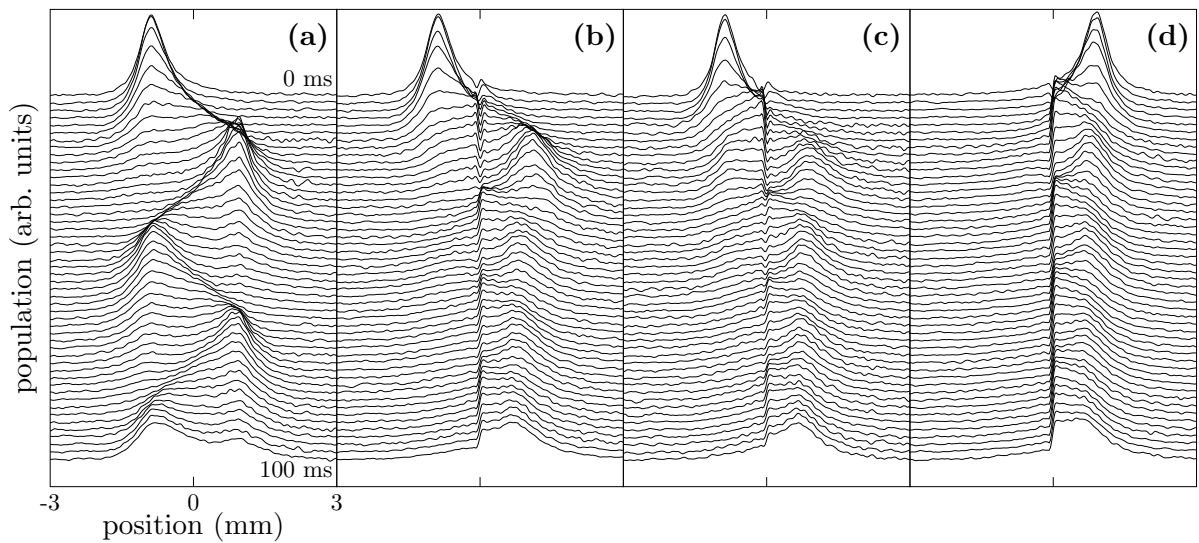}
  \end{center}
  \vspace{-5mm}
  \caption
        {Atom spatial distributions along the dipole trap responding
        to the one-way barrier.  The dipole-trap
        focus and barrier beams are located at the plot centers.
        Each image is an average of 78 repetitions of the experiment.
        Column (a): atoms initially in $F=1$ are dropped from the left
        with no barrier present.
        Column (b): atom initially in $F=1$ are dropped onto the
        barrier from the
        left (transmitting side).
        Column (c): atoms initially in $F=2$ are dropped onto the
        barrier from the
        left.
        Column (d): atoms initially in $F=1$
        are dropped onto the barrier from the right (reflecting side).
      \label{fig:waterfall}}
\end{figure*}

\begin{figure}[tb]
  \begin{center}
  \includegraphics[scale=.94]{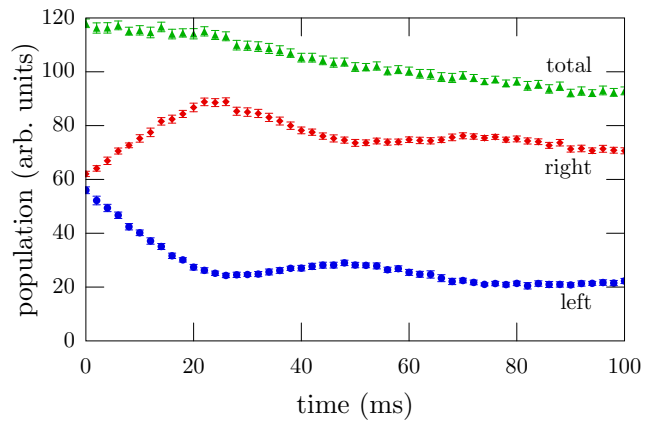}
  \end{center}
  \vspace{-5mm}
  \caption
        {(Color online.)
				Populations on left- and right-hand sides of the one-way barrier,
				where atoms were initially loaded throughout the dipole trap, along
				with the total population.
				Error bars indicate statistical error from $38$ repetitions.
      \label{fig:population}}
\end{figure}

The one-way-barrier beams are separated by $34(1) \ \mathrm{\mu m}$, and their
foci nearly coincide with the focus of the dipole-trap beam, intersecting
it at about $12^\circ$ from the perpendicular to the beam axis (as in
Fig.~\ref{fig:opticalsetup}).
They are asymmetric Gaussian beams, with waists of $11.5(5) \ \mathrm{\mu
m}$ ($13(2) \ \mathrm{\mu m}$) along the dipole-trap axis and $80(7) \
\mathrm{\mu m}$ ($60(7) \ \mathrm{\mu m}$) perpendicular to the dipole-trap
axis for the main (repump) barrier beam.
The repumping barrier beam is derived from the same laser source as the MOT
repump, and is thus also resonant with the MOT repump ($F=1 \rightarrow F'=2$)
transition, with a power of $0.36(4) \ \mathrm{\mu W}$ (as seen by the
atoms).
The main barrier beam has $40(4) \ \mathrm{\mu W}$ of power and is stabilized on
the $^{85}$Rb $F=3 \rightarrow F'=3,4$ crossover dip in the saturated-absorption
spectrum, which is $1.16(6) \ \mathrm{GHz}$ blue of the $^{87}$Rb MOT trapping
transition.
These beams are turned on at time $t=0$ in Figs.~\ref{fig:waterfall}
and \ref{fig:population}.

We use probe absorption for imaging and measuring atom number.
A $45 \ \mathrm{\mu s}$ pulse of probe light resonant with the MOT trapping
transition illuminates the atoms while they are effectively frozen in
place.
The probe beam orientation is nearly perpendicular to the dipole-trap beam,
and the absorption is detected by
a charge-coupled-device (CCD) camera.
To reduce systematic errors due to interference fringes in the images, we
subtract background offsets (as computed from the edge regions of the
images) on a per-column basis.
The distributions in Fig.~\ref{fig:waterfall} are obtained by integrating
the images transverse to the dipole trap.
The spatial resolution is 24.4~$\mu\mathrm{m}$ as set by the CCD pixel
spacing, but the distributions are smoothed slightly for visual clarity.



Our main results are given in Fig.~\ref{fig:waterfall} \cite{EPAPS}, which
shows the atomic evolution in response to the one-way barrier.
As a context for the asymmetric barrier, Fig.~\ref{fig:waterfall}(a)
shows the evolution of atoms beginning to the left of the trap center, in the
\emph{absence} of the barrier.
In this case, the atoms simply oscillate about the center of the dipole
trap, with some breakup of the atomic cloud due to the trap anharmonicity.
Fig.~\ref{fig:waterfall}(b) clearly shows the effects of the one-way barrier.
The atoms begin to the left as in Fig.~\ref{fig:waterfall}(a), on the
transmitting side of the barrier (that is, the main barrier beam is to the
left and the repumping barrier beam is to the right in the images).
We observe the atoms \emph{transmitting} through the barrier on their way
to the right, but when they return, they \emph{reflect} off the right-hand
side of the barrier.
Later, the atomic cloud settles to a steady state to the right of the
barrier.

In Fig.~\ref{fig:waterfall}(b), atoms started out in the transmitting $F=1$
ground level.
But what happens if we repeat the experiment with atoms in the ``wrong,''
\emph{reflecting} ($F=2$) ground level?  This situation is shown in
Fig.~\ref{fig:waterfall}(c), where we see that as expected, many
atoms initially bounce off the barrier, even though they are on the
transmitting side.
However, we see that later on, the atoms \emph{still} manage to make it
through the barrier, without much extra loss.
The atoms can do this because we have chosen the main barrier beam to be
more nearly resonant with the $F=2 \rightarrow F'$ transitions than with the $F=1
\rightarrow F'$ transitions.
This beam thus tends to optically pump atoms into the transmitting
$F=1$ level.
As discussed below, we chose the barrier detuning to have this property,
which also optimizes the performance of the barrier in the presence of
spontaneous scattering of barrier photons.

If we drop the atoms from the right-hand (reflecting) side of the barrier
as in Fig.~\ref{fig:waterfall}(d), we see the atoms \emph{reflect}
from the barrier.
The reflection is clean, even though we have started the atoms in the
``wrong'' $F=1$ ground state, because the repumping barrier beam pumps them
to $F=2$ before they encounter the main barrier beam.
Of course, we observe similarly good reflections
when starting the atoms on the right-hand side in the ``proper'' $F=2$ state.

Lifetimes on the right-hand side of the barrier typically range from $300 \
\mathrm{ms}$ to $500 \ \mathrm{ms}$, depending on the temperature of the
atoms.
At longer times (after warmer atoms are lost), the lifetimes
increase.
Fig.~\ref{fig:waterfall} (d), at longer time scales, had a lifetime of
$>800 \ \mathrm{ms}$.
Losses are mainly due to scattering (and thus heating) from the barrier
beams, both while passing through the main barrier beam and reflecting from
it.

The loss of atoms and optical pumping due to the main barrier beam point to
the main technical challenge associated with implementing the one-way
barrier with $^{87}$Rb atoms: the $6.8 \ \mathrm{GHz}$ ground-state hyperfine
splitting limits the detunings of the main barrier beam to relatively small
values.
The atoms thus scatter more photons than necessary as they cross the
barrier, as we now illustrate.
Let our main barrier beam have a peak potential $V_0<0$ (for atoms in the
transmitting state) and full width at half maximum $w$.
For large detunings $\Delta\gg\Gamma$, where $\Gamma$ is the excited-state
decay rate, the peak scattering rate is given by
$R_\mathrm{sc} = \left|\Gamma V_0/\hbar \Delta\right|$.
Suppose the atoms hit the barrier with kinetic energy $\eta|V_0|$,
determined by the initial atomic displacement.
The total number of scattering events is on the order of
$N_\mathrm{sc}\sim R_\mathrm{sc} w/v =
 w\Gamma\sqrt{m|V_0|}/\hbar|\Delta|\sqrt{2\eta}$,
where $v=\sqrt{2\eta|V_0|/m}$ is the speed of the atoms, assuming it is
constant.
Inserting numbers for our experiment, in particular using $w=2\ln(2) \times
11.5 \ \mathrm{\mu m}$, $\eta=2.2$, and $V_0 = -h \cdot 0.96(10) \
\mathrm{MHz}$, we find $\sim\! 0.7$ scattering events for each
transmission, just due to the main barrier beam.
Note that the unequal detunings imply $\eta < 1$ for the reflecting state,
as required for reflection.
Simulations that account for changes in speed and state estimate
$N_\mathrm{sc} \sim 1$ for transmission.
When we add in the overlapping repump beam, $N_\mathrm{sc}$ increases to
around $3$ during transmission and around $10$ for a single reflection.

An obvious strategy is detuning the barrier between the two
transitions and keeping the overlap between the two barrier beams minimal to
reduce scattering.
This fails because there is no barrier height for which scattering during
\emph{both} transmission and reflection have sufficiently low scattering
rates that the atomic states could be expected to not change.
Our current setup---where the main barrier beam pumps atoms into the
transmitting state---employs some overlap between beams 
to help the atoms reflect properly.
Heating from scattering is worse with this setup, but potentially
catastrophic state changes are minimized.
Substantial deviations from the parameters used here generally resulted in
reduced performance of the barrier in terms of heating and losses.
We will discuss these issues in depth in a future publication
\cite{Thorn08}.

Scattering from the main barrier beam also explains what is happening in
Fig.~\ref{fig:waterfall}(c).
Atoms reflecting off the transmitting side of the barrier are
preferentially pumped to $F=1$, which allows them to transmit.
This process involves extra scattering because the atoms that transmit were
most likely scattered to the $F=1$ state while turning around, and thus
were moving slowly.
This gives more time for scattering to occur.
The expected increase in heating is manifest as an increased loss of atoms.

As a further test of the barrier, we loaded atoms away from the dipole-trap
center for $110 \ \mathrm{ms}$, which uniformly and symmetrically filled
the dipole trap with about $9 \times 10^4$ atoms.
We then activated the barrier with a lower power of $18(2) \
\mathrm{\mu W}$ after letting the atoms equilibrate in the dipole trap for
$200 \ \mathrm{ms}$.
Fig.~\ref{fig:population} shows the population on each side of the
barrier after it is activated \cite{EPAPS}.
Like the trapdoor operated by Maxwell's demon, atoms start on either
side of the barrier, which ``opens'' to let atoms travel to the
right, and ``closes'' to block atoms from traveling to the left.
As time progresses, the atoms end up mostly on the reflecting
side.
Small oscillations due to residual collective motion in the dipole trap are
also visible in this plot.
Clearly atoms are moving from the left-hand side to the right-hand side and
not the reverse, compressing the physical volume in which the atoms reside.
The spatial compression is countered somewhat by spontaneous-scattering
heating in the presence of
the barrier, and we observe an increase of the atomic phase-space density
of $7(2)\%$.
Better compression is likely achievable; however, our present setup
optimizes the barrier asymmetry rather than compression.


In summary, we have demonstrated an all-optical asymmetric potential barrier 
capable of increasing overall phase space density of a sample of
neutral alkali atoms. 


This research was supported by the National Science Foundation,
under project PHY-0547926.




\end{document}